\documentclass[aps,pra,twocolumn,amsfonts,floatfix,superscriptaddress,showpacs]
{revtex4}

\usepackage{dcolumn}
\usepackage{bm}

\usepackage{amssymb}          
\usepackage{graphicx}         
\usepackage[latin1]{inputenc} 
\usepackage{amsmath,amsfonts}          
\usepackage{color}
\usepackage{graphicx} 
\usepackage{fancyhdr} 
\usepackage{listings} 
\usepackage{syntonly} 
\usepackage{psfrag}
\usepackage{multirow}
\usepackage{pgf,pgfsys,pgffor}
\usepackage{pgfplots}
\usepackage{pgfplotstable}
\usepackage{tikz}
\usetikzlibrary{intersections,arrows,positioning,patterns}
\usepackage{cases}
\usepackage{hyperref}
\usepackage{filecontents}
\usepackage{flushend}

\newtheorem{prop}{Proposition}

\newtheorem{cor}{Corollary}

\newcommand{\proof}{\noindent {\bf Proof. }}

\newcommand{\X}{x}			
\newcommand{\Y}{y}			
\newcommand{\x}{x}			
\newcommand{\y}{y}			
\newcommand{\p}{p}			
\newcommand{\sinp}{\rho}	
\newcommand{\sout}{\tau}	
\newcommand{\Hilbert}{\mathcal{H}} 		
\newcommand{\D}{\mathcal{D}}		
\newcommand{\povm}{\Pi}			
\newcommand{\Id}{I_\Hilbert}			
\newcommand{\Od}{0_\Hilbert}			
\newcommand{\plo}{\p_{1|0}}			
\newcommand{\pol}{\p_{0|1}}			
\newcommand{\pll}{\p_{1|1}}			
\newcommand{\poo}{\p_{0|0}}			
\newcommand{\Prob}[1]{\textrm{P} \left[ #1 \right]} 
\newcommand{\I}{H}			
\newcommand{\entropy}{H}			
\newcommand{\Pc}{\textrm{P}_c} 
\newcommand{\Pe}{\textrm{P}_e} 
\newcommand{\capacity}{C}			
\newcommand{\vinp}{\vec \rho}		
\newcommand{\pauli}{\vec \sigma}		
\newcommand{\vout}{\vec \tau}		
\newcommand{\vproj}{\vec \pi}		
\newcommand{\U}{U}		
\renewcommand{\S}{\Lambda}
\newcommand{\V}{V}
\newcommand{\Ell}{E}			
\newcommand{\Sph}{S}			
\newcommand{\trsf}{A}		
\newcommand{\shift}{\vec b}		
\newcommand{\xSys}{\textrm{x}}
\newcommand{\ySys}{\textrm{y}}
\newcommand{\zSys}{\textrm{z}}
\newcommand{\bx}{b_\xSys}
\newcommand{\by}{b_\ySys}
\newcommand{\bz}{b_\zSys}
\newcommand{\Ch}{\mathcal{E}}			
\newcommand{\te}{\theta}			
\newcommand{\ps}{\psi}				
\newcommand{\al}{\alpha}				
\newcommand{\be}{\beta}				
\newcommand{\reg}{\mathcal{V}}	
\newcommand{\diff}{\vec d}		
\newcommand{\hbin}{h_2} 
\newcommand{\diag}{\mathop{\rm diag}\nolimits}
\newcommand{\rad}{R}
\newcommand{\radDelta}{\sqrt{S}}
\newcommand{\generatingV}{\mathcal{G}_\reg}
\newcommand{\borderV}{\mathcal{B}_\reg}
\newcommand{\maximalV}{\mathcal{M}_\reg}
\newcommand{\parallelogramV}{\mathcal{P}}
\newcommand{\frontier}{\mathcal{F}}

\newcommand{\tr}[1]{\textrm{tr} \left( {#1} \right)}

\newtheorem{Observation}{Observation}
\newcommand{\bo}{\begin{Observation}}
\newcommand{\eo}{\end{Observation}}

\definecolor{blueN}{RGB}{0,41,85}

\newcommand{\Hi}{\mathcal{H}}

\newcommand{\beq}{\begin{equation}}
\newcommand{\eeq}{\end{equation}}
\newcommand{\beqa}{\begin{eqnarray}}
\newcommand{\eeqa}{\end{eqnarray}}
\newcommand{\beqan}{\begin{eqnarray*}}
\newcommand{\eeqan}{\end{eqnarray*}}

\newcommand{\qed}{\hfill $\Box$ \vskip 2ex}

\definecolor{nblue}{rgb}{0.3,0.3,1.0}
\definecolor{ngreen}{rgb}{0.2,0.7,0.2}
\definecolor{nred}{rgb}{0.9,0.1,0}

\colorlet{NGREEN}{ngreen} 
\colorlet{NBLUE}{nblue}
\colorlet{VIOLET}{violet}
\colorlet{NRED}{nred}
\colorlet{BLACK}{black}

\begin{document}

\title{Optimal Binary Codes and Measurements for\\ Classical Communication over Qubit Channels}

\author{Nicola Dalla Pozza}
\email{nicola.dallapozza@dei.unipd.it}
\affiliation{Dipartimento di Ingegneria dell'Informazione,
Universit\`a di Padova, via Gradenigo 6/B, 35131 Padova, Italy}
\author{Nicola Laurenti}
\email{nicola.laurenti@dei.unipd.it}
\affiliation{Dipartimento di Ingegneria dell'Informazione,
Universit\`a di Padova, via Gradenigo 6/B, 35131 Padova, Italy}
\author{Francesco Ticozzi}
\email{ticozzi@dei.unipd.it}
\affiliation{Dipartimento di Ingegneria dell'Informazione,
Universit\`a di Padova, via Gradenigo 6/B, 35131 Padova, Italy}
\affiliation{\mbox{Department of Physics and Astronomy,
Dartmouth College, 6127 Wilder Laboratory, Hanover, NH 03755, USA}}

\date{\today}

\begin{abstract}

We propose constructive approaches for the optimization of binary classical communication over a general noisy qubit quantum channel, for both the error probability and the classical capacity functionals. After showing that the optimal measurements are always associated to orthogonal projections, we construct a parametrization of the achievable transition probabilities via the coherence vector representation. We are then able to rewrite the problem in a form that can be solved by standard, efficient numerical algorithms and provides insights on the form of the solutions. 

\end{abstract}


\maketitle

\section{Introduction}

Every communication system relies on a physical layer into which encode information for delivery. The role of the quantum properties of a media in communication protocols was first investigated by the pioneering work by Helstrom \cite{Helstrom1976}, and the study has subsequently evolved into a rich branch of quantum information, see e.g. \cite{Eldar2001, Giovannetti2004}. Most notably, the results of quantum information theory have provided rigorous definitions and results for the (classical and quantum) capacity of quantum channels, see e.g. \cite{Nielsen2004} for an introductory review of early results, and  \cite{Holevo2012} for more recent developments.

In this work, we consider the problem of transmitting classical information over a quantum channel that is not ideal, namely, it is described by a Completely-Positive, Trace-Preserving (CPTP) map \cite{Kraus1983, Nielsen2004}. We aim at finding {\em optimal  input states and output measurements} with respect to some performance index.  We focus on the \emph{binary case}, namely where two ``symbol'' states can be transmitted and two ``detection'' measurement outcomes are considered. An effectively implementable solution of this problem, beside having an obvious relevance {\em per se}, would be also instrumental to real-time optimization of communications over time-varying channels and the improvement of the key generation rate of well established quantum cryptography protocols \cite{Bennett1984,Bennett1992}.

In the literature, two different functionals are typically used to evaluate the quality of a classical digital communication system: symbol {\em error probability} and {\em channel capacity}. 

Adopting the first functional for classical communication \emph{over quantum channels} leads to a problem that is closely related to optimal discrimination. The quantum binary discrimination problem, namely the problem of distinguishing two \emph{given} quantum states with maximal probability, has been addressed by Helstrom, and the form of the optimal measurement operators, as well as the maximal probability of correct  discrimination, have been found analytically for every pair of input states \cite{Helstrom1976}.

Finding the optimal measurement for the discrimination problem is equivalent to the optimization of the receiver for a quantum binary \emph{ideal} communication channel with respect to the error probability.
When a non-ideal channel is considered, and the input states are \emph{fixed}, the optimal measurement operators are the projectors that solve the discrimination problem for the corresponding output states (see e.g. \cite{Nielsen2004}). When a subsequent optimization with respect to the input states is aimed, the number of variables involved in the optimization procedure is still considerable.

Therefore, we solve the problem of optimization with respect to both the states and measurements by deriving necessary conditions for the optimality and by proposing an efficient numerical procedure.


On the other hand, the \emph{capacity of a classical binary channel} represents the maximum amount of information that can be reliably sent from the transmitter to the receiver per use of the channel when only two symbols are employed. It corresponds to the maximum of the mutual information between input and output, computed over all possible a priori distribution and coding of the source. 

We here assume that source bits are encoded into pairs of quantum states, and this encoding as well as the decoding protocol are memoryless so that the cascade of the encoder/quantum channel/decoder is equivalent to a classical binary memoryless channel. The channel therefore is completely characterized by the transition probabilities. 
We then consider the classical capacity of the binary channel we obtain.

Note that this in general different from the \emph{Shannon} capacity $\capacity_{Shan}$, the \emph{one-letter} capacity $\capacity_\chi$ and the \emph{full} capacity $\capacity$ of the \emph{qubit} channel, that are the maximum amount of information that can be sent through the quantum channel with respectively separable quantum states and separable measurement operators, separable input states and joint (possibly entangled) measurement operators, and possibly non-separable states and joint measurements on multiple outputs \cite{Holevo2012}. 
In fact, for a qubit channel, to achieve these capacities it may be necessary to employ up to 4 states \cite{Davies1978}, while we limit ourselves to a pair, and employing separable transmitted states and measurement operators. Qubit channels that achieve capacity using 2 to 4 states have been studied in \cite{Berry2005,Fujiwara1998,King2002,Hayashi2005}, focusing on the situations where the Holevo bound is saturated. Moreover, the additivity property, that allows to identify the full capacity with the one-letter capacity, has been verified in the case of unital \cite{King2002unital} and depolarizing \cite{King2003} qubit channels. In these particular cases, an analytical solution of the maximization problem for $\capacity=\capacity_\chi$ is possible, while a general solution is unknown.

While we are bound to obtain suboptimal performances, the binary assumption allows us to devise a constructive procedure 
by leveraging on necessary conditions on the optimality of the input states and measurement operators \emph{given an arbitrary channel}, and simplify the numerical procedure for the maximization of the binary capacity.

In seeking the joint optimization of the input states and measurement operators for a given channel, we extend the analysis in \cite{Elron2007} to noisy, not necessarily unital channels, by using linear--algebraic and geometric tools similar to those used in \cite{King2001}.
Our approach to the optimization problem allows to develop some insight on the {\em family of classical channels}, represented by their transition probabilities, that can be obtained by properly engineering input states and output measurements for a given quantum channel. We shall show that, for most channels, the optimization with respect to either functionals lead to very similar solutions: however, it is indeed possible to find particular channels for which this is not true. This behavior is related to the curvature of the border of the region of the admissible transition probabilities, as it will be illustrated in some examples.

This paper is organized as follows. 
In Section \ref{system}, we briefly review the communication setup used and the notation. 
In Section \ref{partial} we introduce partial orderings for binary channels and we prove that the optimal measurements for the communication performance indexes are always associated with a pair of orthogonal projectors.
Section \ref{geometric} reviews the coherence vector representation of qubits, shows that the optimal pair of input states are orthogonal and defines the region of admissible transition probabilities.
Section \ref{optimization} is devoted to the optimization of the chosen performance index.
In Section \ref{examples} we give some examples of binary channels and evaluate the corresponding performance indexes.
Section \ref{conclusions} summarizes our results and concludes the paper.

\section{\label{system}Communication Setup}

Let ${\cal M}_2$ denote the set of $2\times 2$ complex matrices. If $X\in{\cal M}_2,$ $X^\dag$ indicates the transpose conjugate of $X.$ A qubit is a two-level system, with associated Hilbert space $\Hi \sim{} \mathbb{C}^2.$ Quantum states are associated to density  matrices, $\D(\Hilbert)=\{\rho\in{\cal M}_2|\rho=\rho^\dag\geq 0,\,\tr{\rho}=1\}.$ Physically admissible maps from states to states are represented by Completely-Positive Trace-Preserving (CPTP) linear maps $\Ch:\D(\Hilbert)\rightarrow \D(\Hilbert).$

\begin{figure}[h]
\begin{center}
\medskip
\includegraphics{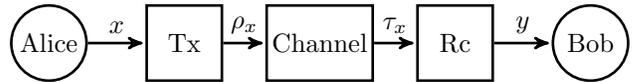}
\end{center}
\caption{Binary Communication Scenario, with $\x, \y \in \{0,1\}.$}
\label{fig:scenario}
\end{figure}

We consider the communication scenario depicted in Figure \ref{fig:scenario}: a sender, Alice, aims to transmit classical bits over a known noisy qubit channel $\Ch$ to a receiver, Bob. Alice is modeled as a random source of bits $\x \in \{0,1\}$, that selects a quantum state $\sinp_\x,\ \x =0,1$ according to the value of $\x$, and sends it over the channel. The media is described by a CPTP map $\Ch$ on $\D(\Hilbert)$. At the receiver, Bob has to reconstruct the transmitted bit/state by performing a measurement on the unknown incoming state, $\sout_x:=\Ch(\sinp_\x)$.

A general binary measurement for a qubit channel on $\Hilbert$ can be described by a Positive-Operator Valued Measure (POVM) $\{\povm_0,\povm_1\},$ where $\povm_0+\povm_1=I$ \cite{Nielsen2004}. The probability of obtaining $\y \in\{0,1\}$ as an outcome of a measurement on a system in the state $\sout$ is then computed as $\tr{\povm_\y\sout}.$ Without loss of generality, we assume that if $\povm_\y$ is measured, than Bob decides that the input was $\y.$
Transition probabilities between the transmitted bit $\x$ and measurement outcome $\y$ can be calculated as
\beqan
\plo & := & \Prob{\y=1|\x=0}= \tr{\povm_1 \sout_0}\\
\poo & := & \Prob{\y=0|\x=0}= 1-\plo=\tr{\povm_0 \sout_0}\\
\pol & := & \Prob{\y=0|\x=1}= \tr{\povm_0 \sout_1}\\
\pll & := & \Prob{\y=1|\x=1}= 1-\pol=\tr{\povm_1 \sout_1}.
\eeqan
We denote by ${\cal C}$ the classical binary channel characterized by transition probabilities $p_{y|x}$. 
When it is evident from the context what are the input states and the measurements that are used, we refer to ${\cal C}$ as the {\em classical} channel associated to $\Ch.$

We assume that Alice and Bob can agree on the protocol to use, i.e. jointly decide the quantum states $\{\sinp_0,\ \sinp_1\}$, the probability distribution $\p_\x$ to use for the bit source and the measurements $\{\povm_0,\povm_1\}$ that are employed in the receiver. This  can be done in an optimal way with respect to a performance index for the binary communication channel. We shall consider two indices, the {\em bit error probability} and the {\em classical channel capacity}. The error probability and its counterpart, the probability of correct decision, are defined respectively as:
\beqa
\Pe & =& \plo \ \p_0 + \pol \ \p_1, \label{Pe} \\
\Pc & =& \poo \ \p_0 + \pll \ \p_1 = 1-\Pe. \label{Pc}
\eeqa
For fixed input states $\{\sinp_\x\}$ and measurements $\{\povm_\y\}$, the classical channel capacity is defined as \cite{Nielsen2004}:
\beq
\capacity_{bin} = \operatorname*{max}_{\p_\x} \ \I(\X;\Y),
\label{cap3}
\eeq
where the mutual information
\beq
\I(\X;\Y)= \sum_{\x,\y}\p_{\y|\x} \p_\x \left(\log \p_{\y|\x}-\log \sum_{\x'}\p_{\y|\x'}\p_{\x'}\right)
\label{mutualInformation}
\eeq
depends on the {\em a priori} probability $\p_\x$.

As we have already anticipated, we achieve the maximization of the functionals leveraging on necessary conditions on the optimality of the measurement operators and on the quantum states. For this reason, the optimization process cannot be split into an inner optimization with respect to the former and an outer optimization with respect to the latter, although in the case of error probability we make use of the Helstrom result \cite{Nielsen2004} on the measurement optimization. 

\begin{figure}
	\centering
	\includegraphics{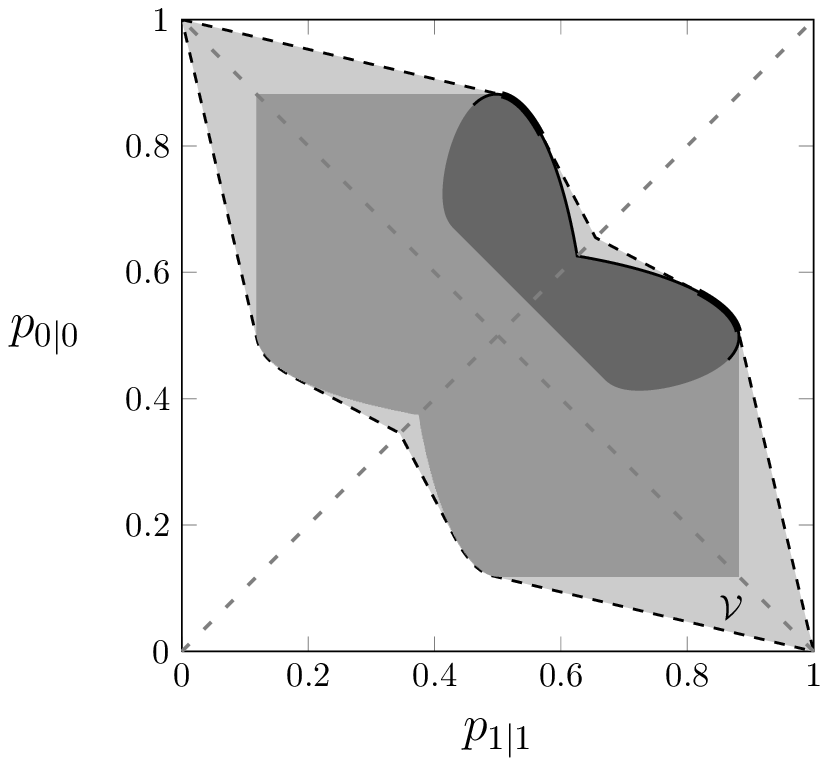}
	\caption{\label{figureReg} Region $\reg$ of admissible transition probabilities, for $\trsf = \diag([.1,.4,.1]),\	\shift = [.23,.32,.05]$, filled with \protect\tikz \protect\filldraw[line width=1pt,fill=black!20!white,dashed] (0ex,0ex) rectangle (3ex,1.7ex);~. The binary channels included in $\reg$ are obtained by POVMs and arbitrary quantum states.
	In \protect\tikz \protect\filldraw[line width=1pt,fill=black!40!white,draw=none] (0ex,0ex) rectangle (3ex,1.7ex);~, the set of binary channels with rank 1 projector measurement operators  and arbitrary quantum states.
	Filled with \protect\tikz \protect\filldraw[line width=1pt,fill=black!60!white,draw=none] (0ex,0ex) rectangle (3ex,1.7ex);~, the set of binary channels with projectors and optimal quantum states given by equation \eqref{normal}.
	The black dashed line \protect\tikz \protect\draw[line width=1pt,no markers, draw=black, dashed] (0ex,0ex) (0ex,.3ex)--(3ex,.3ex); represents $\frontier(\reg)$, the frontier of $\reg$. 
	The thin black line \protect\tikz \protect\draw[line width=1pt,no markers, draw=black] (0ex,0ex) (0ex,.3ex)--(3ex,.3ex); represents $\generatingV$, defined in \eqref{def:generatingV}, while in thick black line \protect\tikz \protect\draw[line width=2pt,no markers, draw=black] (0ex,0ex) (0ex,.3ex)--(3ex,.3ex); it is plotted the set of maximal binary channels $\maximalV$, defined by  the intersection in eq. \eqref{def:maximalV}. 
	In gray dashed lines it is represented the symmetry axes of $\reg$.
	}
\end{figure}

Figure \ref{figureReg} highlights the reduction procedure that occurs including the necessary optimality conditions on the 12 independent variables (see the coherence vector representation in Section \ref{geometric}) describing the input quantum states and measurement operators involved in the optimization process.

From the outside, inward, we show the region of admissible transition probability obtained with arbitrary POVM and arbitrary quantum states,
that with rank 1 measurement projectors and arbitrary quantum states,
that with projectors and optimized quantum states,
and the set of ``maximal'' binary channel, where to seek for the transition probabilities that maximize either the probability of correct decision or the classical binary capacity.

\section{\label{partial}Partial orderings for classical binary channels}
\begin{figure}
	\centering
	\includegraphics{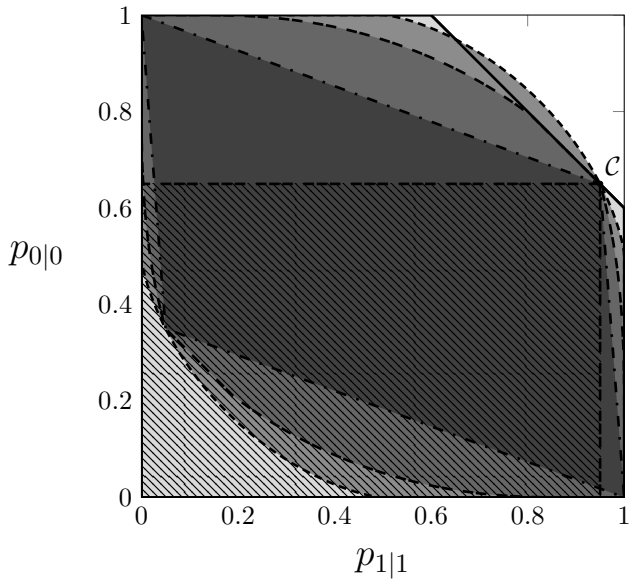}
	\caption{\label{fig:chord}Illustration of the partial ordering with respect to the binary channel $\mathcal C$, characterized by its correct transition probabilities $\pll~=~0.95,\ \poo=0.65$.}
			\begin{tabular}{c|c|c}
		area & border line & sets of channels \\
		\hline 
		\rule{0pt}{5.5ex}
		\begin{tikzpicture}[x=4ex,y=4ex]
		\filldraw [fill=white,draw=black] (0,0) rectangle (1,1); 
		\filldraw [fill=black!15!white,draw=black,line width=1pt] (0,0) -- (0,1) -- (.6,1) -- (1,.6) -- (1,0) -- (0,0) ;
		\end{tikzpicture}
		& \protect\tikz \protect\draw[line width=1pt,draw=black] (0ex,0ex) (0ex,1.5ex) -- (4ex,1.5ex); 
		& \shortstack{with higher error rate than $\mathcal C$ \\ (with equally likely symbols)} \\
		\rule{0pt}{5.5ex}
		\begin{tikzpicture}[x=4ex,y=4ex]
		\filldraw [fill=white,draw=black] (0,0) rectangle (1,1); 
		\filldraw [fill=black!45!white,line width=1pt,draw=black] (1,0) -- (1,0.5) arc (0:90:0.5) -- (0,1) --(0,0.5) arc (180:270:0.5) -- (1,0);
		\end{tikzpicture}
		& \protect\tikz \protect\draw[line width=1pt,draw=black,densely dashed] (0ex,0ex) (0ex,1.5ex) --  (1ex,1.5ex); 
		& with lower capacity than $\mathcal~C$ \\
		\rule{0pt}{5.5ex}
		\begin{tikzpicture}[x=4ex,y=4ex]
		\filldraw [fill=white,draw=black] (0,0) rectangle (1,1); 
		\filldraw [fill=black!60!white,line width=1pt,draw=black] (1,0) arc (0:90:1) arc (180:270:1);
		\end{tikzpicture}
		& \protect\tikz \protect\draw[line width=1pt,draw=black,dash pattern= on 4pt off 1pt on 4pt off 4 pt] (0ex,0ex) (0ex,1.5ex) -- (3ex,1.5ex); 
		& less capable than $\mathcal C$ \\
		\rule{0pt}{5.5ex}
		\begin{tikzpicture}[x=4ex,y=4ex]
		\filldraw [fill=white,draw=black] (0,0) rectangle (1,1); 
		\filldraw [fill=black!75!white,draw=black,line width=1pt] (0,1) -- (0.85,0.65) -- (1,0) -- (0.15, 0.35) -- (0,1);
		\end{tikzpicture}
		& \protect\tikz \protect\draw[line width=1pt,draw=black,dash pattern= on 4pt off 4pt on 1pt off 4 pt] (0ex,0ex) (0ex,1.5ex) -- (5ex,1.5ex); 
		& \shortstack[c]{stochastically degraded \\ with respect to $\mathcal C$} \\
		\rule{0pt}{5.5ex}
		\begin{tikzpicture}[x=4ex,y=4ex]
		\filldraw [fill=white,draw=black] (0,0) rectangle (1,1); 
		\filldraw [pattern=north west lines,line width=1pt] (0,0) rectangle (0.85,0.65);
		\end{tikzpicture}
		& \protect\tikz \protect\draw[draw=black,line width=1pt,dash pattern= on 4pt off 1pt on 4pt off 1pt on 4pt off 4 pt] (0ex,0ex) (0ex,1.5ex) -- (4ex,1.5ex); 
		& dominated by $\mathcal C$
	\end{tabular}	
\end{figure}

A binary memoryless channel $\mathcal C$ can be uniquely represented by its transition probability matrix 
\beq
T_{\mathcal C} = \left[\begin{array}{cc} \poo & \pol \\ \plo & \pll \end{array}\right]
\label{transitionProbabilityMatrix}
\eeq
or, more compactly, by the pair of correct transition probabilities $(\pll,\poo)$. 
In classical Information Theory literature, the problem of comparing discrete channels in terms of their transition probabilities has been studied extensively \cite{Shannon1958,Abramson1960,Helgert1967}. We are interested in the following (partial) orderings for binary memoryless channels
\begin{description}
\item[Product Ordering:] in the standard product ordering, a channel $\mathcal C'$ is dominated by another channel $\mathcal C$ if $\poo' \leq \poo$ and $\pll' \leq \pll$.
\item[Stochastic Degradedness \cite{Korner1977}:] a channel $\mathcal C'$ is stochastically degraded with respect to another channel $\mathcal C$ if $\mathcal C'$ is equivalent to the cascade of $\mathcal C$, followed by a further channel $\mathcal C''$, that is $T_{\mathcal C'} = T_{\mathcal C''} T_{\mathcal C}$.
\item[Capability Ordering \cite{Korner1977,VanDijk1997}:] a channel $\mathcal C'$ is said to be less capable than another channel $\mathcal C$ if, for any input $\x$, by denoting with $\y,\y'$ the corresponding outputs from $\mathcal C,\mathcal C'$, respectively, we have $\I(\x;\y') \leq \I(\x;\y)$.
\end{description}
By using the representation of channels as points $(\pll,\poo)$ in the unit square, Figure \ref{fig:chord} shows the sets of channels that are dominated by, stochastically degraded with respect to, or less capable than channel $\mathcal C$.

The following relations hold: i) if $\mathcal C'$ is dominated by $\mathcal C$ in the product ordering, then the probability of correct decision is not higher in $\mathcal C'$ than in $\mathcal C$, with any input distribution; ii) if $\mathcal C'$ is stochastically degraded with respect to $\mathcal C$ then $\mathcal C'$ is also less capable than $\mathcal C$ by the data processing inequality \cite{Cover2006}; iii) if $\mathcal C'$ is less capable than $\mathcal C$, then $\mathcal C'$ has a lower capacity than $\mathcal C$; however, all the above inclusions are strict, as the converse statements are false, in general.
Moreover, if we restrict our attention, without loss of generality, to the channels for which $\poo'+\pll' \geq 1$, we also have: iv) if $\mathcal C'$ is dominated by $\mathcal C$ in the product ordering, then $\mathcal C'$ is also stochastically degraded with respect to $\mathcal C$; v) if $\mathcal C'$ is stochastically degraded with respect to $\mathcal C$, then the probability of correct decision with equally likely inputs is not higher in $\mathcal C'$ than in $\mathcal C$.

Due to the above implications, while we aim at optimizing channels in terms of correct decision probability or capacity, we will make use of both the product ordering and the stochastic degradedness notions, as they are simpler to assess in terms of the channel transition probabilities. 

The following statement shows that for binary channels orthogonal projectors are always optimal over all POVMs in terms of stochastic degradedness. 

\begin{prop} 
\label{prop1}
For any binary channel ${\cal C'},$ associated to input states $\{\sinp_0,\sinp_1\}$ and POVM elements $\{\povm_0',\povm_1'\}$, there exists a binary channel ${\cal C}$ associated to the same input states and to a pair of {\em orthogonal projections} $\{\povm_0,\povm_1\}$	such that ${\cal C'}$ is stochastically degraded with respect to ${\cal C}$.
\end{prop}

\proof 
Since $\povm'_0+\povm'_1=I,$ it is easy to show that $\povm_0'$ and $\povm_1'$ must be simultaneously diagonalizable \cite{Albertini2011}. Choosing an appropriate basis we can thus write them as:
\[\povm'_0=\diag(q_a,q_b),\ \povm'_1=\diag(1-q_a,1-q_b)\]
with $0\leq q_i \leq 1,\ i=a,b$.
In the same basis, we can represent the channel output states $\sout_x=\Ch(\sinp_\x)$ as 
\beq
\sout_0 = \left[ \begin{array}{cc}
\lambda_0 & \ast \\
\ast & 1-\lambda_0
\end{array}\right],
\quad
\sout_1 = \left[ \begin{array}{cc}
\lambda_1 & \ast \\
\ast & 1-\lambda_1
\end{array}\right]
\eeq
where entries denoted by $\ast$ are irrelevant for our analysis.
The binary channel resulting by $\{\povm_0',\povm_1'\}$ is associated with the transition probability matrix 
\beq
T_{\mathcal C'} = \left[\begin{array}{cc} 
q_b + (q_a-q_b)\lambda_0 & q_b + (q_a-q_b)\lambda_1 \\
(1-q_b) + (q_b-q_a)\lambda_0 & (1-q_b) + (q_b-q_a)\lambda_1
\end{array}\right].
\eeq
Now, let us consider the following projectors:
$$
\povm_0=\diag(1,0),\  \povm_1=\diag(0,1)
$$
and observe that the corresponding transition probability matrix is
\beq
T_{\mathcal C} = \left[\begin{array}{cc} 
\lambda_0 & \lambda_1\\ 
1-\lambda_0 & (1-\lambda_1) 
\end{array}\right].
\eeq
If we define the transition probability matrix
\beq
T_{\mathcal C''} = \left[\begin{array}{cc} 
q_a & q_b \\ 
1-q_a & 1-q_b
\end{array}\right]
\eeq 
it is trivial to see that $T_{\mathcal C'} = T_{\mathcal C''} T_{\mathcal C}$, resulting in the stochastic degradedness definition for $\mathcal C'$ with respect to the channel $\mathcal C$.\qed

A similar conclusion, i.e. that orthogonal rank-1 measurement operators are optimal for the classical channel capacity functional, can be obtained as a corollary of the results presented in \cite{Davies1978}. However, the approach we follow shows that for qubit channels this is a consequence of a stronger ordering property, namely stochastic degradedness.

In the case of the probability of correct decision, a similar condition holds,
as stated in the following proposition. This result can already be found in \cite{Helstrom1976}, but here we provide an alternative proof in the context of channel ordering on the base of Proposition \ref{prop1}.  

\begin{prop}
\label{prop2}
For any binary channel ${\cal C'},$ associated to input states $(\sinp_0,\sinp_1)$ and POVM elements $(\povm_0',\povm_1')$,  and any input distribution $(\p_0,\ \p_1)$ there exists a binary channel ${\cal C}$ associated to the same input states and to a pair of orthogonal projections $(\povm_0,\povm_1)$ that has probability of correct decision better or equal than ${\cal C}$.
\end{prop}
\proof 
Consider the pair $(\povm_0,\povm_1)$ derived from $(\povm'_0,\povm'_1)$ as in Proposition \ref{prop1}, yielding the transition probabilities $(\pll,\poo)$. To this, add the trivial projector pairs $(\Id,\Od)$ and $(\Od,\Id)$ which yield transition probabilities $(1,0)$ and $(0,1)$, respectively. 
The original $(\pll',\poo')$ lie in the triangle of vertices $\{(1,0),\ (\pll,\poo),\ (0,1) \}$ and since the probability of correct decision \eqref{Pc} is a linear function of the transition probabilities, the proof follows from the fact that the extremal values of a linear function on a polytope are always found on vertices. \qed
 
By combining Propositions \ref{prop1} and \ref{prop2} with implications ii) and iii) about channel orderings, it is easy to derive the following result:
\begin{cor}
The optimal measurements for either functional are always associated to a pair of orthogonal projectors.
\end{cor}

It was already recognized in \cite{Tomassoni2008} that the optimal
measurement operators for the binary discrimination problem with respect
to the error probability and the mutual information are projectors. The
authors also showed that if the output of the channel $\sout_\x$ are
pure states, the optimal projectors  for the error probability coincide
with the ones for the mutual information. In our work we prove the
optimality of projectors in the context of stochastic degradedness,
which leads to the same result, but is more general and establishes a
direct link to classical channel hierarchy. Furthermore, in the next
sections we shall show that the optimal projectors for the two
functionals need not be the same, in general.

\section{\label{geometric}Coherence Vector Representation and Geometric Picture }

In order to determine the optimal probabilities, it is crucial to understand how the channel transforms the transmitted quantum states, and then determine the achievable transition probabilities. Hence, we first focus on the characterization of the region of achievable transition probabilities within the unit square.

For our purpose, it is convenient to use a particular choice of basis for representing $2\times2$ complex matrices, associated to the unitary, self-adjoint operators $\{I,\sigma_x,\sigma_y,\sigma_z\},$ where $\sigma_i$ are the Pauli operators, also called \emph{coherence vector representation}.
Input and output quantum states can then be represented as 
\beq
\sinp_\x=\frac{1}{2}(\Id +\vinp_\x \cdot \pauli),\quad \sout_\x=\frac{1}{2}(\Id+\vout_\x \cdot \pauli)
\eeq
where for any $\vec{v} \in \mathbb{R}^3$, $\vec{v} \cdot \pauli$ is the shorthand notation for the linear combination of Pauli matrices
\beq
\vec{v} \cdot \pauli = \vec{v}(1) \ \sigma_x + \vec{v}(2) \ \sigma_y + \vec{v}(3) \ \sigma_y .
\eeq
All the valid (i.e., unit-trace, positive-semidefinite) states are associated to coherence vectors $\vec{v}$ in the unit (Bloch's) sphere, with pure states lying on the surface \cite{Nielsen2004}. 

As we showed in the previous Section, the optimal choice of  measurements for Bob is represented by a pair of projectors $\{\povm_0,\povm_1\}$. Leaving aside the trivial projector pairs $\{\Id, \Od\}$, we consider rank-$1$ projectors that admit coherence representation
\beq
\povm_\y=\frac{1}{2}(\Id +\vproj_\y \cdot \pauli)
\eeq
with $\vproj_\y$ lying on the sphere surface, where the completeness relation $\povm_0+\povm_1=\Id$ implies the constraint 
\beq
\vproj_{0} = -\vproj_{1}.
\label{completeness}
\eeq

The qubit channel is described by a TPCP map $\Ch$ acting on a two level system $\Hilbert$.
In coherence vector representation, any TPCP map has an affine form \cite{Ruskai2002}
\beq 
\label{affine}
\vout_\x=\trsf \ \vinp_\x + \shift,
\eeq
with $\trsf$ being a $3 \times 3$ real matrix associated to a contraction (not necessarily strict), and $\shift$ a vector in the unit ball corresponding to the image of the completely mixed state through the channel. Geometrically, this means that the image $\Ell$ of the Bloch sphere $\Sph$ is an ellipsoid: if $\trsf = \U \  \S \ \V^T$ is the SVD decomposition of $\trsf$, $\Sph$ is first rotated by $\V$, squeezed along its axes by $\S$, rotated again by $\U$ and then shifted by $\shift$.
Note that, as it has been pointed out in \cite{Ruskai2002}, not all maps of the form \eqref{affine} mapping the Bloch sphere into itself yield a physical (i.e. CP) channel. Since in our work the channel is assumed to be physical and known, this is not a concern.

Any channel can be reduced, via change of basis for $\vinp$ and $\vout$, to the case of a diagonal $\trsf$,
\beq
\trsf = \S = \textrm{diag}(a,b,c) .
\label{DiagonalTrsf}
\eeq
In fact, we can define 
\beqa
{\vec \psi} & =& \V^T \vinp \label{Vt} \\
{\vec \phi} & = & \U^T \vout \\
{\vec \xi}  & = & \U^T \shift
\label{basisChange}
\eeqa
so that \eqref{affine} becomes
\beq
{\vec \phi} = \S {\vec \psi} + \xi.
\eeq
From ${\vec \psi}$, we can get $\vinp$ of the original coordinate system by inversion of \eqref{Vt}. In the coordinate system where $\trsf$ is diagonal, the ellipsoid $\Ell$ has axes parallel to those of the standard $(x,y,z)$ coordinate system.

\subsection{Optimization of input states for given projectors}

Transition probabilities can be written in terms of the coherent representation of states and projectors by using inner product in $\mathbb{R}^3$
\beq
\begin{aligned}
\pll & =  \frac{\displaystyle 1+\vproj_1 \cdot \vout_1}{2} , \\
\poo & =  \frac{\displaystyle  1+ \vproj_0 \cdot \vout_0}{2} = \frac{\displaystyle 1-\vproj_1 \cdot \vout_0}{2} .
\end{aligned}
\label{trans}
\eeq

If, as is often the case,  $a,b,c <1$ 
no point of the ellipsoid lies on the sphere surface. Consequently it is not possible to have $\vproj_\y \cdot \vout_\x = 1,\ \y,\x=0,1$ and the region $\reg$ of admissible transition probability is strictly contained in the unit square.

\begin{figure}
\centering
	\includegraphics{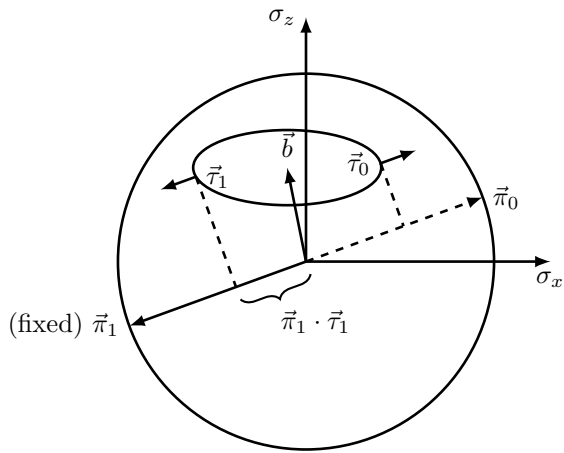}
	\caption{\label{fig:projOttimi} Inner product between $\vproj_1$ and $\vout_1$, with the Bloch sphere projected onto the $\{\sigma_x,\sigma_z\}$ plane. For a \emph{fixed} $\vproj_1$, the point $\vout_1$ on the ellipsoid that maximizes $\vproj_1 \cdot \vout_1$ has normal vector to the surface which is parallel to $\vproj_1$ (here depicted unnormalized). }
\end{figure}

\begin{prop}
\label{prop3}
Let $(\vproj_0,\vproj_1)$ be the coherence vector representation of a pair of orthogonal projectors. Denote with $\vout_0, \vout_1$ the points on the surface of $\Ell$ where the normal vector to the surface is parallel to $\vproj_0,\vproj_1$ respectively. Then the binary channel associated with $(\vout_0, \vout_1)$ dominates with respect to product ordering all the binary channel associated with other states pairs in the ellipsoid.
\end{prop}
\proof Consider $\vproj_1$ fixed as shown in Figure \ref{fig:projOttimi}. By standard results in constrained optimization \cite{Boyd2004}, the output vector $\vout_1$ that corresponds to the  maximum $\pll$ must identify a point on the surface of $\Ell$, with normal vector parallel to $\vproj_1$. In fact, if we consider a plane normal to $\vproj_1$, all the points in the intersection with the ellipsoid correspond to vectors $\vout$ with equal inner product with $\vproj_1$. Hence, they give the same transition probability $\pll$. Among the planes that are orthogonal to $\vproj_1$, the one that maximizes the inner product is thus the plane tangent to $\Ell$ and closer to $\vproj_1$. Analogously for $\vout_0$. \qed

Recalling that $\vproj_0 = -\vproj_1$, the vector $\vout_0$ that maximizes $\vproj_0 \cdot \vout_0$ is then the point on the surface of $\Ell$ with normal vector $-\vproj_1$, and is the ``antipodal'' point of $\vout_1$ in the ellipsoid, that is 
\beq
\vout_0 + \vout_1 = 2 \shift.
\label{antipodal}
\eeq

Note that the antipodal condition \eqref{antipodal} on $\vout_0,\ \vout_1$ implies that the corresponding input vectors $\vinp_0,\ \vinp_1$ are also antipodal, on the Bloch sphere, meaning that the quantum states $\sinp_0,\ \sinp_1$ are orthogonal. This is true in both the case of matrix $\trsf$ with full or deficient rank, since the inverse image of $\vout_\x$, with $\vout_\x$ on the surface of the ellipsoid (full rank $\trsf$) or on the border of the disk (rank deficient $\trsf$), is unique and lie on the surface of the Bloch sphere. 
Since \eqref{antipodal} is a necessary condition for the optimization, we have the following result
\begin{cor}
If the projectors $\{\povm_0, \povm_1\}$ have rank 1, the optimal quantum states to transmit for either functionals are orthogonal. 
\end{cor}

The fact that an orthogonal alphabet of quantum states is a necessary condition for optimality for the classical channel capacity functional can already be found in \cite{Fuchs1997}, however here we examine in depth the optimization, deriving the relations between the optimal transmitted quantum state and the optimal receiver measurements.

We now derive a parametrization of $\vout_0,\vout_1$ in an appropriate coordinate system.
This allows us to impose the condition on the gradients which is necessary for optimality, and write a simplified form of the transition probabilities.

The origin of the coordinate system is the center of $\Sph$ and its axes are parallel to those of $\Ell$. We can parametrize the point $\vproj_1$ on the surface of $\Sph$ by the angles $\al \in [-\frac{\pi}{2},\frac{\pi}{2}], \be \in [0,2\pi)$ and the point $\vout_1$ on the surface of $\Ell$ with $\te \in [-\frac{\pi}{2},\frac{\pi}{2}], \ps \in [0,2\pi)$, 
\beq
\vproj_1 = \left[ \begin{array}{c}
\cos \al \cos \be \\
\cos \al \sin \be \\
\sin \al
\end{array} \right], \ 
\vout_1 = \left[ \begin{array}{c}
a \cos \te \cos \ps +b_x\\
b \cos \te \sin \ps +b_y\\
c \sin \te +b_z
\end{array} \right].
\label{PiTauDef}
\eeq
In order to find the desired $\vout_1$, in Appendix \ref{gradient} we show that setting its gradient equal to $\vproj_1$ leads to the following conditions,
\beq
\begin{array}{c}
a \tan \ps  = b \tan \be, \\
\tan \te \sqrt{a^2 \cos^2 \be + b^2 \sin^2 \be}   = c \tan \al,
\end{array}
\label{normal}
\eeq
and the resulting inner product, necessary to evaluate \eqref{trans}, is
\begin{align}
&\vproj_1 \cdot \vout_1  =  \left( a \cos \te \cos \ps + b_x \right)  \cos \al \cos \be \notag \\
&\quad   + \left( b \cos \te \sin \ps + b_y \right)\cos \al \sin \be + \left( c \sin \te +b_z \right) \sin \al \notag \\
&\quad = \sqrt{ a^2 \cos^2 \al \cos^2 \be +  b^2 \cos^2 \al \sin^2 \be+ c^2\sin^2 \al}  \notag\\
&\qquad \quad  +  b_x \cos \al \cos \be + b_y \cos \al \sin \be + b_z \sin \al.
\label{inner1}
\end{align}
Similarly, by \eqref{antipodal},
\beqa
\vproj_0 \cdot \vout_0  & = & -\vproj_1 \cdot (2 \shift - \vout_1) = \vproj_1 \cdot \vout_1 - 2 \vproj_1 \cdot \shift \notag\\
& = & \sqrt{a^2 \cos^2 \al \cos^2 \be +  b^2 \cos^2 \al \sin^2 \be+ c^2\sin^2 \al} \notag \\
& & -  b_x \cos \al \cos \be - b_y \cos \al \sin \be - b_z \sin \al.
\label{inner0}
\eeqa

\subsection{Region of achievable transition probabilities}


In this Section we characterize the set of transition probabilities obtained as the vector $\vproj_1$ changes on the surface of the Bloch sphere, employing necessary conditions \eqref{normal} for the optimal quantum states $(\vout_0,\vout_1)$ out of the channel.

Let us define the set $\reg$ in the unit square containing transition probabilities pair $(\pll',\poo')$ corresponding to the binary channel ${\cal C'}$ given by generic POVM and generic quantum states. Region $\reg$ shows evident properties of symmetry, with respect to the bisecting line $\pll = \poo$ and the anti-bisecting line $\pll + \poo = 1$.

In fact, given a pair $(\pll',\poo') \in \reg $ obtained from the POVM pair $(\povm_0', \povm_1')$ with the received quantum states $(\sout_0, \sout_1)$, swapping the measurement operators pair and the transmitted quantum states we obtain respectively transition probabilities $(1-\pll',1-\poo')$ and $(1-\poo',1-\pll')$, that are the points symmetrical to $(\pll',\poo')$ with respect the central point $(0.5,0.5)$ and with respect to line $\pll + \poo = 1$. Combining both swaps, we get the symmetry with respect to the bisecting line.
\begin{prop}
\label{prop4}
For all $k\in [-\|\shift\|,\|\shift\|]$, there exist a $\vproj_1$ with $\vproj_1 \cdot \shift = k$ such that the channel ${\cal C}$ with $\vproj_0=-\vproj_1$ and $(\vout_0, \vout_1)$ as given by \eqref{antipodal} and \eqref{normal},  dominates all the channels ${\cal C'}$ similarly associated to any $\vproj_1'$ such that $\vproj_1'~\cdot~\shift~=~k$. Such $\vproj_1$ is given by
\beq
\vproj_1 = \displaystyle \operatorname*{argmax}_{\vproj_1' \in \Sph,\ \vproj_1' \cdot \shift = k} \quad \operatorname*{max}_{\vout_1 \in  \Ell} \quad  \vproj_1' \cdot \vout_1,
\label{edgeProblem}
\eeq
where the solution of the inner maximization problem is given by \eqref{normal}.
\end{prop} 
\proof From \eqref{antipodal}, we can rewrite transition probability $\poo$ \eqref{trans} as
\beq
\poo = \frac{1+\vproj_1 \cdot \vout_1}{2} - \vproj_1 \cdot \shift 
\eeq
so that with $\vproj_1 \cdot \shift $ fixed, \eqref{edgeProblem} maximizes both $\pll$ and $\poo$. \qed


On the basis of propositions \ref{prop1}--\ref{prop4}, all the channels that are \emph{maximal} \footnote{In a set with a partial ordering, an element is said to be \emph{maximal} if it does not precede any other element.} with respect to the stochastic degradedness ordering satisfy equation \eqref{edgeProblem}. 
The optimal channel with respect to either functional should be sought within the set of maximal channels indicated by the thick black line in Figure \eqref{figureReg}. 
We thus propose the following procedure for an efficient evaluation of $\reg$: 

\begin{itemize}
\item[1.] For each value  $k=\vproj_1 \cdot \shift,\ k \in [0,\ \| \shift\|_2]$, solve \eqref{edgeProblem}. This problem is equivalent to a quadratic problem with quadratic constraints (see Appendix \ref{qpqc}), that has no closed form solution but that can be easily solved via standard numerical methods \cite{Boyd2004}. 

\item[2.] Consider the set of channels associated with the previous solutions, and mirror this set with respect to the bisecting line, obtaining the set of channels
\begin{align}
\generatingV= & \{(\pll,\poo) \ | \ \vproj_1 \textrm{ solution of }\eqref{edgeProblem}, \notag \\ 
 & \qquad \qquad  k \in [-\| \shift\|_2,\ \| \shift\|_2]\}. 
\label{def:generatingV}
\end{align}

\item[3.] Consider the parallelogram $\parallelogramV (\bar{\p})$ parametrized in $\bar{\p}=(\pll, \poo) \in \generatingV$  with vertices $\{(0,1),\bar{\p},(1,0),(1,1)-\bar{\p}\}$. The region $\reg$ is given by the union of the parallelograms for $\bar{\p} \in \generatingV$,
\beq
 \reg = \bigcup_{\bar{\p} \in \generatingV} \parallelogramV (\bar{\p}),
\label{regioneV}
\eeq
and it is depicted in Figure \ref{figureReg} filled with \protect\tikz \protect\filldraw[line width=1pt,fill=black!20!white,dashed] (0ex,0ex) rectangle (3ex,1.7ex);~.
\end{itemize}

We can define the upper border of $\reg$ from its frontier $\frontier(\reg)$ and the set $\mathcal{S}_{+}=\{(\pll, \poo): \pll+\poo \geq 1\}$, as in 
\beq
\borderV = \frontier(\reg) \cap \mathcal{S}_{+}.
\eeq
The border $\borderV$ allows to define the set of maximal channels of $\Ch$ as
\beq
\maximalV = \generatingV \cap \borderV.
\label{def:maximalV}
\eeq
Figure \eqref{figureReg} shows an example of the sets of binary channel that can be obtained from a quantum channel $\Ch$, with different measurement operators (POVM, projectors and optimal projectors) and different quantum states (arbitrary and optimal ones). On the border of the region, the frontier of $\reg$ in dashed line, the set of channels $\generatingV$ in thin black line and the set of maximal channels $\maximalV$ in thick black line.


\section{\label{optimization}Optimization and Numerical Methods}

\subsection{Probability of correct decision}

In the case of probability of correct decision, we consider the a priori symbol probabilities $\p_0,\p_1$ as given. 
By exploiting the geometric representation of the previous section, we can rewrite the problem so that we can obtain a solution via standard numerical methods, as well as  an insightful geometrical picture.
Combining the definition of $\Pc$ with the relation of completeness, we get
\beq
\Pc = (1-\p_1) + \tr{\povm_1 (\p_1 \sout_1 - \p_0 \sout_0)} .
\label{Pc2b}
\eeq
Following Helstrom \cite{Helstrom1976}, in order to find optimal solution for the problem of quantum binary discrimination it is convenient to introduce the difference operator $\Delta = \p_1 \sout_1 - \p_0 \sout_0$. We now use the coherence vector representation for $\povm_1,\ \sout_1, \ \sout_0$ to get
$$
\Delta= \frac{1}{2} \left( (1-2\p_0)\Id + \diff \cdot \pauli \right),$$ $$ \diff = \p_1 \vout_1 -\p_0 \vout_0 = \vout_1 - 2 \p_0 \shift :=(d_x,d_y,d_z)^T
$$
where the antipodal condition \eqref{antipodal} has been used. From \eqref{Pc2b}, we see that the optimal $\povm_1$ is the projection on the eigenspace of $\Delta$  associated to positive eigenvalues. The eigenvalues of $(1-2\p_0)I+ \diff \cdot \sigma$ are 
\beq
\begin{aligned}
	\lambda_0 &=\frac{1-2\p_0-\| \diff \|_2}{2}, \\ 
	\lambda_1 &=\frac{1-2\p_0+\| \diff \|_2}{2}. 
\end{aligned}
\label{eig}
\eeq
Depending upon $\p_0,\ \vout_1,\ \vout_0$, the eigenvalues may be both positive, both negative or opposite in sign. 

In particular, given a qubit channel whose ellipsoid $\Ell$ is strictly contained in the Bloch sphere, such that $\| \vout_1 \|_2<1$ and the ellipsoid radii $a,b,c < 1$, there exists for certain a $\p_0^\ast$ such that either $\lambda_0,\ \lambda_1 >0 \ \ \forall \p_0 \in [0,\p_0^\ast)$ or $\lambda_0,\ \lambda_1 <0 \ \ \forall \p_0 \in (\p_0^\ast,1]$.
Consequently, $\povm_1$ is respectively the identity or the null observable on $\Hilbert$, and it results 
\beq
\Pc = \max \left \{ \p_0, \p_1 \right \}
\label{Pc2c}
\eeq
such that performing a measurement does not increase the probability of correct discrimination with respect to our a priori information. 


In the case $|1-2\p_0|<\sqrt{d_x^2+d_y^2+d_z^2}$, instead, we have $\lambda_{0}>0$ and $\lambda_{1}<0$ and $\povm_1$ is a rank 1 projector as in Proposition \ref{prop2}. We rewrite
\beq
\Pc=\frac{1}{2} + \| \p_1 \sout_1 - \p_0 \sout_0 \|_1 = \frac{(1+ \vproj_1 \cdot \diff)}{2}
\label{Pc2d}
\eeq
This expression gives an immediate meaning to the optimal $\vproj_1$, which must be parallel to $\diff$, and highlights that, in order to maximize $\Pc$, $\diff$ must be taken of the maximum possible length. 

Substituting \eqref{antipodal} in \eqref{Pc2d}, we obtain the expression
\beq
\Pc = \frac{1}{2} \left( 1 + \vproj_1 \cdot \vout_1  - 2 \p_0 \vproj_1 \cdot \shift \right), 
\eeq
which shows two opposing terms, $\vproj_1 \cdot \vout_1$ to be maximized and $\vproj_1 \cdot \shift$ to be minimized. These two terms are related in \eqref{edgeProblem}, where for each value of the latter term the former one is maximized. In principle, we could use \eqref{edgeProblem} and compute a table with the pairs $(\vproj_1 \cdot \vout_1, \vproj_1 \cdot \shift)$ and then maximize the probability of correct decision. However, this is not necessary, since it suffices to solve the quadratic optimization problem
\beq
\diff_{opt} := \operatorname*{max}_{\vout_1 \in \Ell} \| \vout_1 - 2 \p_0 \shift \|_2 
\label{probEq}
\eeq
where $\| \cdot \|_2$ is the Euclidean norm. Numerical methods for convex optimization are well known \cite{Boyd2004} and can be employed to solve the quadratic problem \eqref{probEq} with quadratic constraints. 

Finally, depending on the norm of $\diff_{opt}$, we optimize the measurement operator, which is $\vproj_1$ parallel to $\diff_{opt}$ in the case of $|1-2\p_0|<\sqrt{d_x^2+d_y^2+d_z^2}$, leading to two rank~1 projective operators, or the POVM corresponding to the identity and null operator to associate to the most probable and less probable quantum state.
In the end we get
\beq
\Pc = \max \left \{ \frac{(1+\|\diff_{opt}\|_2)}{2}, \p_0, \p_1 \right \}.
\label{Pc3}
\eeq


\subsection{\label{mutual}Classical binary capacity}

\begin{figure}
	\centering
	\includegraphics{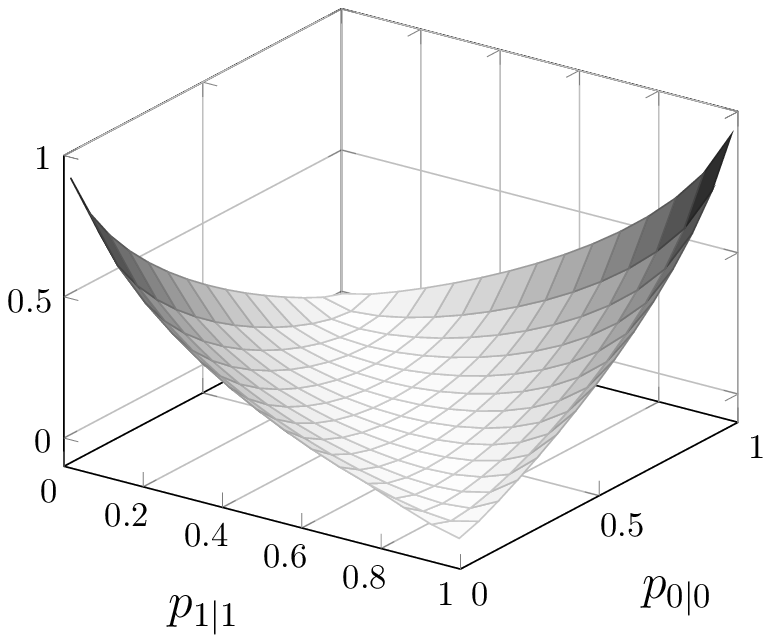}
  \caption{Binary capacity as a function of $\pll$ and $\poo$.}
  \label{fig:capacity}
\end{figure}

Maximization of mutual information \eqref{mutualInformation} requires optimization over many parameters: the a priori probability distribution $\p_\x$, the input states $\sinp_\x$ and the receiver measurement $\povm_\y$. Also, due to nonlinear terms, explicit solutions are difficult to find. Instead, numerical maximization is viable thanks to convexity of the mutual information (see Figure \ref{fig:capacity}). 

From the geometric representation of states and measurement projectors, we can optimize \eqref{mutualInformation} by a search over the region $\reg$. In fact, we can split the maximization problem of \eqref{mutualInformation} into two optimization problems: a inner maximization with respect to the {\em a priori} probability, and an outer maximization with respect to the transition probabilities:
\beq
\capacity_{bin} = \operatorname*{max}_{(\pll,\poo) \in \reg} \ \operatorname*{max}_{\p_\x} \ \I(\X;\Y)
\label{cap4}
\eeq

The inner maximization of \eqref{cap4} has an analytic closed form solution. Consider $(\pll,\,\poo)$ as fixed, 
define the binary entropy
\beq
\hbin(x) = -x \log_2 x -(1-x) \log_2 (1-x).
\label{binaryEntropy}
\eeq
with derivative
\beq
\hbin'(x) = \log_2 \left( \frac{1}{x}-1 \right),
\label{hbinDerivative}
\eeq
and inverse function of the derivative 
\begin{align}
g(x) = (\hbin')^{-1} = \frac{1}{2^x+1}.
\label{inverseHbinDer}
\end{align}
We rewrite the mutual information as
\beqa
 \I(\X;\Y) & =& \entropy (\Y) - \entropy(\Y|\X)  \nonumber\\
 & =& \entropy (\Y)  - \entropy (\Y|\X=1) \ \p_1 - \entropy(\Y|\X = 0) \ \p_0  \nonumber\\
 & =& \hbin(r) - \hbin(\pll) \ \p_1 - \hbin(\poo) \  (1-\p_1), \nonumber \\
  \label{mutualInformation2}
\eeqa
where $r=\Prob{\Y = 0}= \poo (1-\p_1) + (1-\pll) \p_1$.
If we take the derivative of the above with respect to $\p_1,$ we get
\beqan
\frac{\textrm{d} \I(\X;\Y)}{\textrm{d}\p_1}& =& \hbin'(r)\cdot\frac{\textrm{d}r}{\textrm{d}\p_1}-\hbin(\pll)+\hbin(\poo) \\
& =& \hbin'(r)(1-\pll-\poo)-\hbin(\pll)+\hbin(\poo).
\eeqan
By imposing it be equal to zero, we have:
\beq
\hbin'(r) = \frac{\hbin(\pll)-\hbin(\poo)}{1-\pll-\poo}.
\eeq
Hence, by definition
\beq
\begin{array}{rcl}
r & =& g \left( \frac{\displaystyle \hbin(\pll)-\hbin(\poo)}{\displaystyle 1-\pll-\poo} \right) \; , \\
\p_1 & =& \frac{\displaystyle r - \poo}{\displaystyle 1-\pll-\poo} \; ,
\end{array}
\label{p1opt}
\eeq
and maximal mutual information is obtained if we substitute these values in  \eqref{mutualInformation2}.

Outer optimization with respect to $\pll,\poo$ can be performed on $\maximalV$ employing standard tools from constrained optimization. In the light of symmetry considerations, two or four optimal solutions can be found, or even a continuous arc of optimal solutions can be obtained if these points lay on a contour line. 

Since a numerical optimization is required to solve \eqref{edgeProblem} and hence to solve the outer optimization, we can find the points of $\generatingV$ and perform the maximization all together: for a certain $k \in [0,\ \| \shift\|_2]$, first solve \eqref{edgeProblem} to find the optimal $\vproj_1$, get $\vout_1$ by \eqref{normal} and $\vout_0$ by \eqref{antipodal}. From $\vproj_1,\ \vout_1$ and $\vout_0$ obtain $(\pll,\poo)$ by \eqref{trans}. With these transition probabilities, find $\p_1$ from \eqref{p1opt} and get $\I(\X;\Y)$. These steps need to be repeated for a finite set of values $\{k\}$ that discretizes $[0,\ \| \shift\|_2]$, and by direct comparison we can get the maximum $\I_k(\X;\Y)$.

Of course, since a numerical procedure is required, the discretization of the range $[0,\ \| \shift\|_2]$ is necessary in order to calculate and compare the values $\I_k(\X;\Y)$ for different $k$. However, due to smoothness nature of the functional, it is assured that it is possible to find a solution $\I_{\bar{k}}$ arbitrarily close to the true one, i.e.
\beq
\forall \ \epsilon>0 \quad \exists N,\bar{k}>0 \quad \textrm{s.t.} \quad  |\I_{\bar{k}}-\I_{\hat{k}}|< \epsilon 
\eeq
where $\I_{\bar{k}}$ is the maximal mutual information obtained on the $N$--step discretization of the range while $\I_{\hat{k}}$ is the true optimum on $[0,\ \| \shift\|_2]$.

\section{\label{examples}Examples}

The procedure explained in section \ref{geometric} and described in more details in Appendix \ref{qpqc}, allows us to find the region of transition probabilities given a description of the physical channel as in \eqref{affine}. 

Set the numerical routine for the calculation of the border, with a Monte Carlo simulation on the parameter of $\trsf$ and $\shift$, we easily find out different region shapes induced by the channels. Of course, not all the possible choice of entries of $\trsf$ and $\shift$ define a physical channels, so we have to check the necessary and sufficient conditions given in \cite{Ruskai2002}.

\begin{figure}
	\centering
	\includegraphics{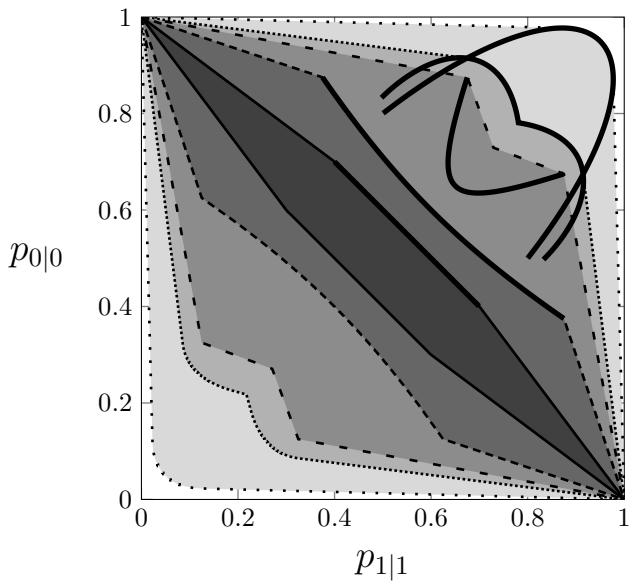}
	\caption{\label{regioni}Different shapes of $\reg$. The corresponding $\generatingV$, defined in \eqref{def:generatingV},  are highlighted in thick black line. In the Table below it is reported the associated ellipsoid parameters. Notice that case 5 presents a border parallel to the anti--bisecting line, leading to a continuum of optimal points with respect to the error probability.}
		\begin{tabular}{c|c|clcr}
		case& $\reg$ & $\trsf$ & & $\shift$ & \\
		\hline
		1 & \protect\tikz \protect\filldraw[line width=1pt,fill=black!15!white,draw=black,loosely dotted] (0ex,0ex) rectangle (5ex,1.7ex);  & $\diag([0.3,0.3,0.9])$ & [0.3,& 0,& 0] \\
		2 & \protect\tikz \protect\filldraw[line width=1pt,fill=black!30!white,draw=black,densely dotted] (0ex,0ex) rectangle (5ex,1.7ex); & $\diag([0.2, 0.1, 0.62])$ & [0.3,& 0,& 0.15]  \\
		3 & \protect\tikz \protect\filldraw[line width=1pt,fill=black!45!white,draw=black,loosely dashed] (0ex,0ex) rectangle (5ex,1.7ex); & $\diag([0.55, 0.3, 0.3])$ & [0.2,& 0,& 0] \\
		4 &\protect\tikz \protect\filldraw[line width=1pt,fill=black!60!white,draw=black,densely dashed] (0ex,0ex) rectangle (5ex,1.7ex); & $\diag([0.25, 0.25, 0.2])$ & [0.5,& 0,& 0]  \\
		5 & \protect\tikz \protect\filldraw[line width=1pt,fill=black!75!white,draw=black,solid] (0ex,0ex) rectangle (5ex,1.7ex); & $\diag([0.1, 0.1, 0.1])$ & [0.3,& 0,& 0] 
	\end{tabular}	
\end{figure}

Figure \ref{regioni} shows different kind of shapes that can be obtained from the channel. The corresponding $\trsf$ and $\shift$ are reported in the caption. Both convex (1,5) and non-convex (2,3,4) regions are possible, with different shapes of border. Since there's no analytical solution for the borders, we cannot find a clear dependence between the ellipsoid parameters and the shape of $\reg$. However, we can develop some intuitions and qualitative analysis on the shape of the region $\reg$ and on the position of the optimal points.

In some particular cases we can interpret the shape of the region in the light of the ellipsoid parameters. For example, in the case of an ellipsoid with equal radii (sphere), we can see that due to the symmetry, the set of the binary channel with rank--1 projectors and arbitrary states is a stripe along the anti--bisecting line, whose thickness depends on the radius and whose length depends on the norm of $\shift$. Another case is when the channel is unital and $\shift=0$. In this case the set of the binary channel with rank--1 projectors becomes a square centered in (0.5,0.5), with the side depending on the longest radius.

As seen in Section \ref{optimization}, the transition probabilities to maximize the probability of correct decision or the mutual information lie on the border of $\reg$. While the contour curves of $\Pc$ are straight lines with slope depending on the a priori probabilities, the contour lines of $\capacity_{bin}$ are bent. In general, the optimal transition probabilities differ depending upon the functional considered. 

In cases 1, 3, 4, 5 depicted in Figure \ref{regioni}, either the solutions coincide, or at least a pair of coinciding solutions exists. Notice that whenever the region $\reg$ presents a border parallel to the anti--bisecting line (as in case 5), the optimization of error probability with equally likely inputs leads to a continuum of solutions on the segment of the border.

\begin{figure}
	\centering
	\includegraphics{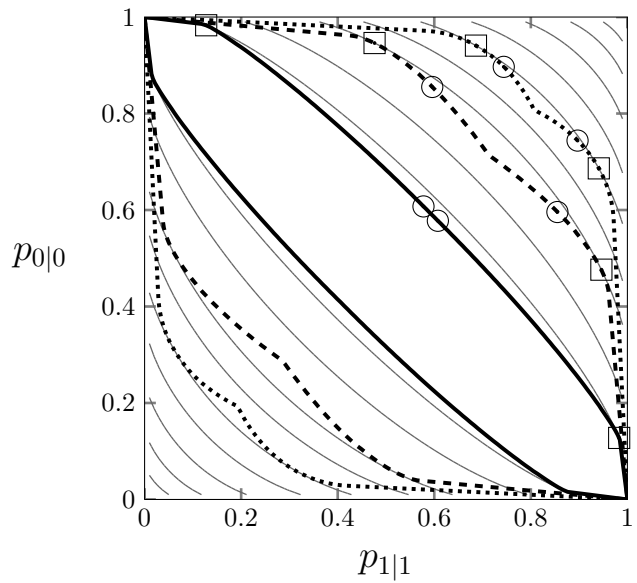}
	\caption{\label{differentPoints} Examples of regions $\reg$ where the point of maximal mutual information \protect\tikz \protect\draw[black,  thin] (0ex,0ex) rectangle (6pt,6pt); does not coincide with the point of maximal probability of correct decision \protect\tikz \protect\draw[black, thin] (0ex,0ex) circle(3pt);. In the background, contour lines of mutual information maximized with respect to $\p_\x$ are depicted in thin gray solid lines. }
	\begin{tabular}{c|ccccc}
		line & $\trsf$ & & $\shift$ & & $\p_1$ \\
		\hline
		\protect\tikz \protect\draw[black,  line width=1.5pt] (0ex,0ex) -- (5ex,0ex); & $\diag([0.14,0.07,0.19])$ & [0.46,& 0.74,& 0.03] & 0.57 \\
		\protect\tikz \protect\draw[black, dashed,  line width=1.5pt] (0ex,0ex) -- (5ex,0ex); & $\diag([0.34, 0.24, 0.45])$ & [-0.42,& -0.27,& -0.26] & 0.55 \\
		\protect\tikz \protect\draw[black, dotted,  line width=1.5pt] (0ex,0ex) -- (5ex,0ex); & $\diag([0.11, 0.64, 0.07])$ & [-0.24,& -0.15,& 0.45] & 0.54
	\end{tabular}	
\end{figure}

The peculiarity of case 2 is further clarified in Figure \ref{differentPoints}, where the region $\reg$ corresponding to three different channels exhibiting an analogous behavior are depicted in solid, dashed and dotted lines, and the optimal points for the different functionals exhibit significant difference. In the background (in thin gray solid lines) the contour lines of the mutual information are drawn to illustrate why the optimal points do not coincide: the local curvature of the border of $\reg$ is lower than the curvature of the mutual information contour line. In doing this comparison, we consider an equal a priori probability for the probability of correct decision. 

This particular situation can arise for both concave and convex regions. However, a convex region with the point of maximal classical capacity along the bisecting line has the same point of maximal probability of correct decision with equal a priori probability.

\section{\label{conclusions}Conclusions}
In the present work we have provided reliable methods to obtain optimal quantum states and measurement that maximize either the probability of correct binary detection or the classical binary capacity for a known arbitrary qubit channel. 

Analytical approaches are in general not viable, and standard numerical optimization methods are unsuccessful because of the non-convexity of the contour surfaces. On the other hand, building on a geometrical representation we can reformulate the problem and resort to a quadratic problem with quadratic constraints that allows for accurate and stable solutions. 

A numerical method is proposed also for the nonlinear channel capacity functional, exploiting a two-step optimization procedure. The inner maximization with respect to the {\em a priori} probability admits a closed-form solution that can be used to simplify optimization with respect to the transition probability. 
In doing so, relying on the classical ordering of communication channels, we proved necessary conditions for the optimality of states and measurements that can be employed to lighten the optimization.
Numerical results as well as  qualitative analysis of the contour plots of the functionals suggest that, even if typically the solutions for the two functionals are very close, considerable differences can emerge in particular cases, depending on the curvature of the boundaries of the admissible transition probabilities. 

While the results can be directly use as presented for the optimization of binary communication, further study is needed to address the advantage of the method with respect to the full capacity channel optimization and to assess the potential of the results in specific realistic scenarios. For instance, we believe that the proposed approach may be useful in designing quantum cryptography protocols over noisy qubit channels to maximize key~rates.


\section*{Acknowledgements}

This work has been carried out within the Strategic-Research-Project QUINTET of the Department of Information Engineering, University of Padova and the Strategic-Research-Project QUANTUMFUTURE of the University of Padova.

\appendix

\section{\label{gradient}}

Consider the coherence vector representation for the quantum states in input and output of the channel $\vinp, \vout$ and for the measurement operators $\vproj$. The affine relation between $\vinp$ and $\vout$,
\beq
\vout = \trsf \vinp + \shift,
\label{affine2}
\eeq
maps the Bloch ball surface associated to
\beq
\vinp^{\ T} \vinp = 1
\eeq
into the ellipsoide $\Ell$ with equation
\beq
(\vout - \shift)^T \trsf^{-T} \trsf^{-1} (\vout - \shift) =1.
\label{Ell}
\eeq

The (unnormalized) normal vector to the surface of $\Ell$ in the point located by $\vout$ can be written as
\beq
{\vec \nabla_{\vout}} = 2\trsf^{-T} \trsf^{-1} (\vout - \shift), \quad \vout \in \Ell.
\eeq
In order to find the point in the ellipsoid with normal vector equal to $\vproj$, we set
\beq
\frac{\trsf^{-T} \trsf^{-1} (\vout - \shift)}{\| \trsf^{-T} \trsf^{-1} (\vout - \shift) \|} = \vproj.
\eeq
After substituting \eqref{affine2}, the expression becomes
\beq
\frac{\trsf^{-T} \vinp}{\| \trsf^{-T} \vinp \|} = \vproj.
\label{gradRho}
\eeq

In particular, by inversion of \eqref{gradRho} we obtain the following relation
\beq
\vinp^{\ T} \vinp = 1= \|\vinp\|^2  = \| \trsf^{-T} \vinp \|^2 \| \trsf^T \vproj \|^2,
\label{norma}
\eeq
and hence 
\beq
\vinp = \| \trsf^{-T} \vinp \| \trsf^T \vproj=\frac{\trsf^T \vproj}{\| \trsf^T \vproj \|},
\eeq
which is equivalent to \eqref{normal}.

In addition, if we evaluate the inner product we get
\beq
\vproj \cdot \vout = \vproj \cdot (\trsf \vinp + \shift) = \vproj \cdot \trsf \vinp + \vproj \cdot \shift
\eeq
and using \eqref{gradRho} and \eqref{norma}, we get
\beq
\vproj \cdot \trsf \vinp = \frac{\vinp^{\ T} \trsf^{-1} \trsf \vinp}{\| \trsf^{-T} \vinp \|} = \frac{\|\vinp\|^2}{\| \trsf^{-T} \vinp \|} = \|\trsf^T \vproj\|.
\eeq
We finally obtain 
\beq
\vproj \cdot \vout = \|\trsf^T \vproj\| + \vproj \cdot \shift,
\eeq
which is the expression used to evaluate \eqref{inner1}.

\section{\label{qpqc}}
Consider the problem \eqref{edgeProblem}, and define the cost function
\beq
f := \vproj_1 \cdot (\vout_1 -\shift) = \vproj_1 \cdot \Delta \vout_1  
\label{funct}
\eeq
and the constraints
\begin{align}
& \vout_1 \in  \Ell, \label{v1}\\
& \vproj_1 \in \Sph, \label{v2}\\
& \vproj_1 \cdot \shift = k, \label{v3}
\end{align}
for $k \in \left[0,\|\shift\|_2\right]$. 
According to definitions \eqref{PiTauDef}, the maximization of \eqref{funct} with constraints \eqref{v1}-\eqref{v3} requires an optimization with respect to four variables, i.e. $\al,\ \be, \ \te$ and $\ps$. 
As already pointed out previously in Section \ref{geometric}, the constraint \eqref{v1} and the geometric interpretation of the optimization allow us to obtain the necessary conditions \eqref{normal}, so that we can substitute $\te,\ \ps$ in terms of $\al,\ \be$, 
to get an expression of the cost function $f$ similar to \eqref{inner1}, i.e.
\beq
\tilde{f} := \sqrt{ a^2 \cos^2 \al \cos^2 \be +  b^2 \cos^2 \al \sin^2 \be+ c^2\sin^2 \al}.
\label{funct1}
\eeq
Alternatively, if we substitute $\al,\ \be$ in terms of $\te,\ \ps$ we obtain
\beq
\tilde{f}' := \frac{abc}{\sqrt{  a^2 b^2 \sin^2 \te +  b^2 c^2 \cos^2 \te \cos^2 \ps  + a^2 c^2 \cos^2 \te \sin^2 \ps }}.
\label{funct2}
\eeq
Depending on the choice of the variables, the optimization problem becomes the maximization or minimization of the square root term in $\tilde{f}$ or $\tilde{f}'$. Also, we can simplify the formulation using $\tilde{f}^2$ or $(\tilde{f}')^2$ as functional, in order to get rid of the square root term.

The constraint \eqref{v2}, that can be rewritten as
\beq
\vproj_1^{\ T} \vproj_1 = 1,
\label{v2b}
\eeq
has intersection with the plane \eqref{v3} that defines a circle on $\Sph$ as region of optimization.

If we consider the points defined by the variables $\al,\be$ in \eqref{PiTauDef}, the cost function $\tilde{f}$ can be interpret as the norm of vector $\vproj_1$ in a non-normal coordinate system, and the problem
\beq
\displaystyle (\vproj_1)_{\textrm{max}}(k) = \operatorname*{argmax}_{\vproj_1^{\ T} \vproj_1 = 1,\ \vproj_1 \cdot \shift = k} \tilde{f}
\label{maxProblem}
\eeq 
becomes a quadratic problem with quadratic constraints. 
The same type of problem can be obtained if we consider the square root term of $\tilde{f}'$ to be minimized.

We develop a reformulation of \eqref{maxProblem} into the problem of finding the farthest point of an ellipse from a given point. We test this approach comparing it with other numerical optimization techniques, such as a ``brute force'' algorithm, that discretizes the ellipse and look for the best $\vproj_1$ in the discretization, and general numerical constrained optimization, that includes the constraints in the functional and finds the maximum with numerical iterative methods. 

First, consider problem \eqref{maxProblem} not as function of variables $\al, \be$ but as function of the coordinates system $\vproj_1 = [\xSys,\ySys,\zSys]^T$ defined by \eqref{PiTauDef}. The cost function of the problem becomes
\beq
\tilde{f}^2 = a^2 \xSys^2 + b^2 \ySys^2 + c^2 \zSys^2,
\eeq
with constraints for a given $k \in \left[0,\|\shift\|_2\right]$
\begin{align}
& \xSys^2 + \ySys^2 + \zSys^2 = 1, \label{v4}\\
& \xSys\ \bx + \ySys\ \by + \zSys\ \bz = k. \label{v5}
\end{align}
The coordinates need first to be normalized \footnote{In the following, the subscript of variables refers to the step in the substitutions.}, with
\begin{align}
\left[\begin{array}{c}
\xSys \\
\ySys \\
\zSys
\end{array}\right]
= 
\left[\begin{array}{ccc}
\frac{1}{a} & 0 & 0\\
0 & \frac{1}{b} & 0 \\
0 & 0 & \frac{1}{c}
\end{array}\right] 
\left[\begin{array}{c}
\xSys_1 \\
\ySys_1 \\
\zSys_1
\end{array}\right] 
:= H_1 \ 
\left[\begin{array}{c}
\xSys_1 \\
\ySys_1 \\
\zSys_1
\end{array}\right],
\label{h1}
\end{align}
and substitute $\zSys_1$ by the constraint \eqref{v5}:
\begin{align}
\left[\begin{array}{c}
\xSys_1 \\
\ySys_1 \\
\zSys_1
\end{array}\right] & = 
\left[\begin{array}{ccc}
1 & 0 & 0\\
0 & 1 & 0 \\
- \frac{c \bx}{a \bz} & - \frac{c \by}{b \bz} & 0
\end{array}\right]
\left[\begin{array}{c}
\xSys_2 \\
\ySys_2 
\end{array}\right]
+
\left[\begin{array}{c}
0 \\
0 \\
\frac{c k}{\bz}
\end{array}\right] \notag \\[10pt]
& := 
H_2\ \left[\begin{array}{c}
\xSys_2 \\
\ySys_2 
\end{array}\right] + t_2.
\label{h2}
\end{align}
We then express the cost $\tilde{f}^2$ and \eqref{v4} as a function of $\xSys_2,\ \ySys_2$. Next, with the change of variables
\beq
\left[\begin{array}{c}
\xSys_2 \\
\ySys_2
\end{array}\right] = \frac{1}{\sqrt{b^2 \bx^2 + a^2 \by^2}}
\left[\begin{array}{cc}
b \bx & a \by \\
a \by & -b \bx
\end{array}\right] \left[\begin{array}{c}
\xSys_3 \\
\ySys_3
\end{array}\right]
:= H3 \left[\begin{array}{c}
\xSys_3 \\
\ySys_3
\end{array}\right],
\label{h3}
\eeq
and
\begin{align}
\left[\begin{array}{c}
\xSys_3 \\
\ySys_3
\end{array}\right] & =
\left[\begin{array}{cc}
\frac{ab}{\rad} & 0 \\
0 & \frac{1}{\bz}
\end{array}\right] \left[\begin{array}{c}
\xSys_4 \\
\ySys_4
\end{array}\right]
+
\left[\begin{array}{c}
\frac{k a b c^2 \sqrt{b^2 \bx^2 + a^2 \by^2}}{\rad^2} \\
0
\end{array}\right] \notag \\[10pt]
 & := H4 \left[\begin{array}{c}
\xSys_4 \\
\ySys_4
\end{array}\right]
+ t_4,
\label{h4}
\end{align}
where $\rad = \sqrt{a^2 b^2 \bz^2 + c^2 b^2 \bx^2 + c^2 a^2 \by^2}$, we obtain a quadratic functional in the canonical form: 
\beq
\tilde{f}^2 = \xSys_3^2 + \ySys_3^2 \ .
\label{functX3}
\eeq
The substitutions \eqref{h1}-\eqref{h4} applied to the constraint \eqref{v4} give a shifted and rotated ellipse. 

We can rotate the coordinate system so that the ellipse has axes parallel to the system's with the substitution
\beq
\left[\begin{array}{c}
\xSys_4 \\
\ySys_4
\end{array}\right] = 
H_5
\left[\begin{array}{c}
\xSys_5 \\
\ySys_5
\end{array}\right],
\eeq
where matrix $H_5$ is: 
\begin{widetext}
\beq
H_5 := 
\left[\begin{array}{cc}
\frac{v^2 - v(\bx^2+\by^2)c^2 + (b^2\bx^2-a^2\by^2)(b^2-a^2)\bz^2 + \radDelta}{n_1} & 
\frac{2\bx \by \bz (b^2-a^2)\sqrt{b^2\bx^2+a^2\by^2}}{n_2} \\[15pt]
\frac{2\bx \by \bz (b^2-a^2)\sqrt{b^2\bx^2+a^2\by^2}}{n_1} & 
\frac{-v^2 + v(\bx^2+\by^2)c^2 - (b^2\bx^2-a^2\by^2)(b^2-a^2)\bz^2 + \radDelta}{n_2}
\end{array}\right]
\label{h5}
\eeq
where $\radDelta = \sqrt{(b^2-c^2)\bx^2+(a^2-c^2)\by^2+(a^2-b^2)\bz^2-4\bx^2\bz^2(a^2-b^2)(b^2-c^2)}$, $v = b^2\bx^2+a^2\by^2$ and $n_1,\ n_2$ are coefficients introduced to normalize the first and second column of $H_5$, respectively. Furthermore, define
\begin{subequations}
\begin{align}
A & = \frac{(a^2+b^2) \bz^2+(b^2+c^2) \bx^2+(a^2+c^2) \by^2+\radDelta}{2 \rad^2}, \label{aConica}\\
B & = 0, \\
C & = \frac{(a^2+b^2) \bz^2+(b^2+c^2) \bx^2+(a^2+c^2) \by^2-\radDelta}{2 \rad^2},\\
D & = \frac{-2 k (b^2-a^2) \bx \by \bz v}{n_1 \rad^2}\left(\bz^2 (a^2 b^2 - c^2 (a^2+b^2)) + \rad^2 - c^4 (\bx^2+\by^2) - c^2 \radDelta \right), \\
E & = \frac{-2 k (b^2-a^2) \bx \by \bz v}{n_2 \rad^2}\left(\bz^2 (a^2 b^2 - c^2 (a^2+b^2)) + \rad^2 - c^4 (\bx^2+\by^2) + c^2 \radDelta \right), \\
F & = \bz^2 \left(\frac{k^2 (a^4 b^4 \bz^2 + b^4 c^4 \bx^2 + a^4 c^4 \by^2)}{\rad^4}-1\right). \label{fConica}
\end{align}
\end{subequations}
\end{widetext}
With this substitutions, the constraint \eqref{v4} is rewritten in the quadratic form
\beq
A\  \xSys_5^2 + B\  \xSys_5 \ySys_5 + C\  \ySys_5^2 + D\  \xSys_5 + E\  \ySys_5 + F = 0,
\label{finEllisse}
\eeq
with coefficients in \eqref{aConica}-\eqref{fConica}. The center and radii of the ellipse result to be
\begin{align}               
\xSys_E &= -\frac{D}{2 A}, \label{xe}\\
y_E &= -\frac{E}{2 C},\\
r_\xSys &= \sqrt{\frac{D^2 C + E^2 A-4 A C F}{4 A^2 C}},\\
r_\ySys &= \sqrt{\frac{D^2 C + E^2 A-4 A C F}{4 A C^2}}. \label{ry}
\end{align}
The problem \eqref{maxProblem} becomes
\begin{align}
\operatorname*{argmax}_{A  \xSys_5^2 + B  \xSys_5 \ySys_5 + C  \ySys_5^2 + D  \xSys_5 + E \ySys_5 + F = 0} \quad \xSys_5^2+\ySys_5^2
\label{ellipseProblem}
\end{align}
that means to find the point on the ellipse described in \eqref{finEllisse} that is farthest form the origin (0,0). 
Problem \eqref{ellipseProblem} still require a numerical algorithm, but this formulation has been tested with respect to other methods mentioned above and results to be the most accurate.

The final vector $\vproj_1$ can be calculated reverting the changes of variables,
\begin{align}
\left[\begin{array}{c}
\xSys_2 \\
\ySys_2
\end{array}\right] & = H_3 \left( H_4 H_5 \left[\begin{array}{c}
\xSys_5 \\
\ySys_5
\end{array}\right] +t_4 \right), \\
\vproj_1 = \left[\begin{array}{c}
\xSys \\
\ySys \\
\zSys
\end{array}\right] &= H_1 \left( H_2 \left[\begin{array}{c}
\xSys_2 \\
\ySys_2 \\
0
\end{array}\right] + t_2 \right).
\end{align}
\bibliographystyle{apsrev}
\bibliography{qbinarychannel}

\end{document}